\documentclass[12pt,twoside]{article}
\usepackage{ae} 
\usepackage[T1]{fontenc}
\usepackage[ansinew]{inputenc}
\usepackage{amsmath}
\usepackage{amssymb}
\usepackage{graphicx}
\numberwithin{equation}{section}
\begin{document}
\bibliographystyle{plain}
\title{\Large \textbf{Comment to “Recent Climate Observations Compared
to Projections” by Rahmstorf et al.}}
\author{\textsc{Gerhard Kramm}\\\\
University of Alaska Fairbanks, Geophysical Institute\\
903 Koyokuk Drive, P.O. Box 757320 Fairbanks, AK 99775-7320, USA\\
Email: kramm@gi.alaska.edu\\
Phone: + 1 907 474 5992}
\date{}
\maketitle
\begin{figure}[t]
\begin{center}
\includegraphics[width=.75\textwidth,height=!]{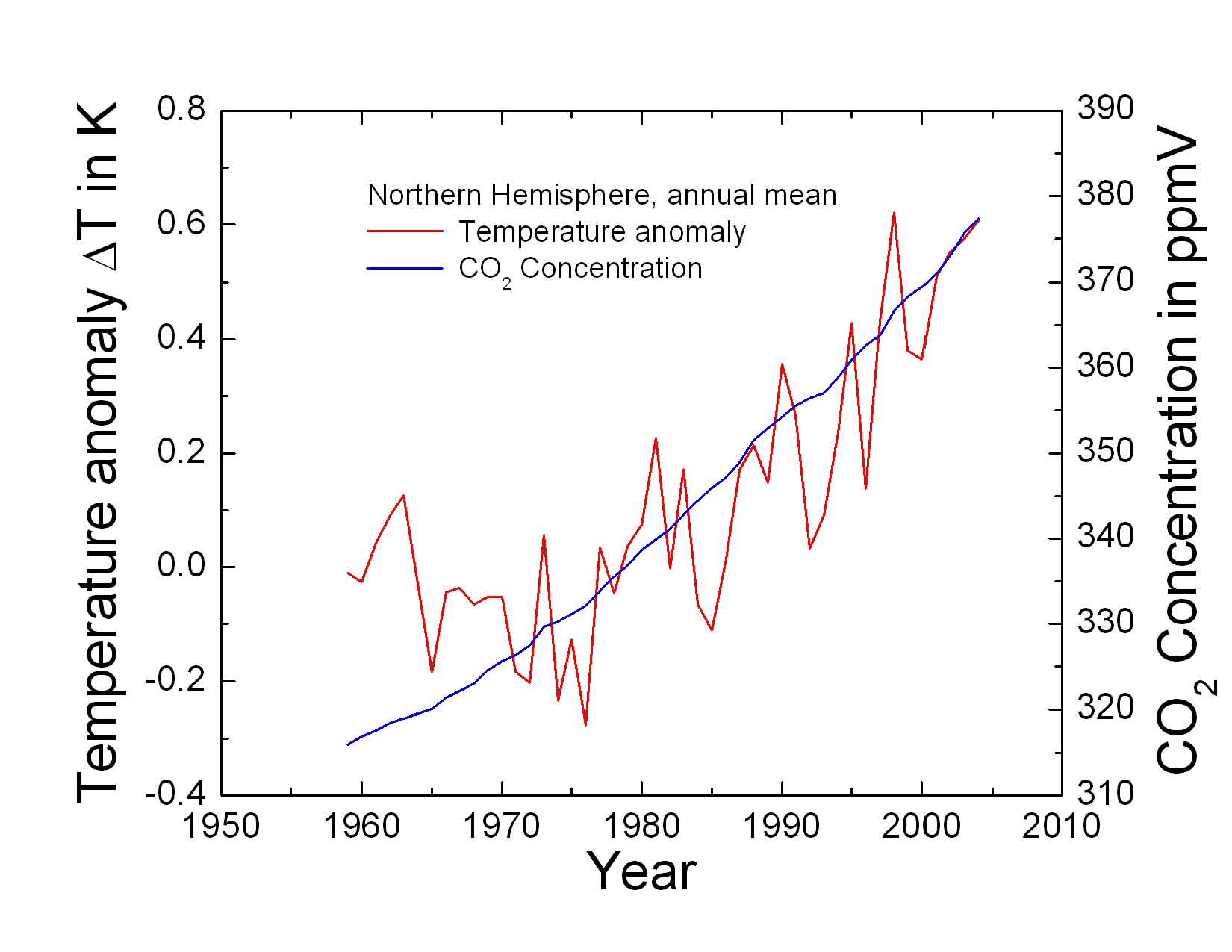}
\end{center}
\begin{center}
\includegraphics[width=.75\textwidth,height=!]{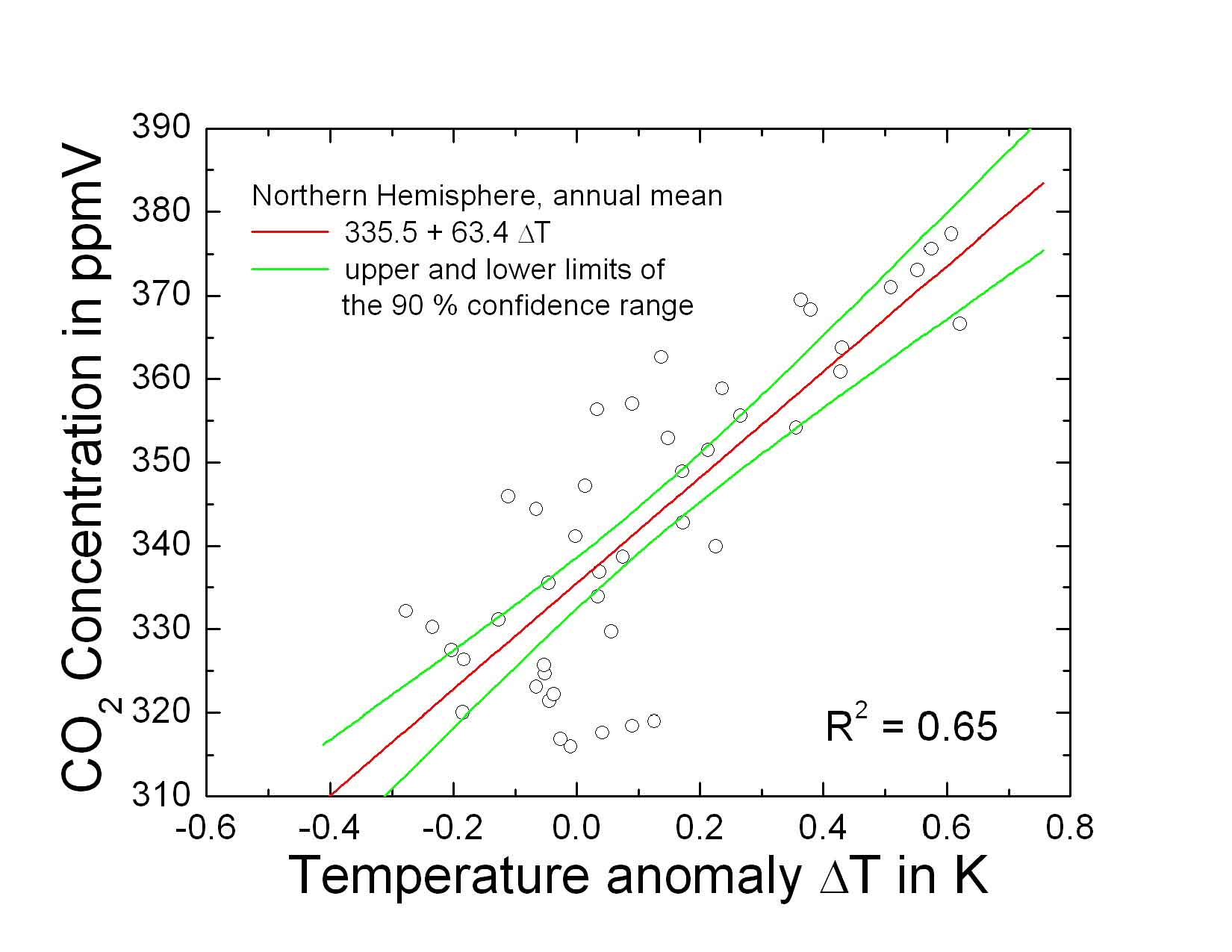}
\end{center}
\caption{Mean annual near surface temperature anomaly and
atmospheric $CO_2$ concentration vs. time for the period 1958 to
2004 (upper part). Also shown are the correlation between the
corresponding values of the temperature anomaly and the $CO_2$
concentration (lower part).} \label{Figure_1}
\end{figure}
\begin{figure}[t]
\begin{center}
\includegraphics[width=.75\textwidth,height=!]{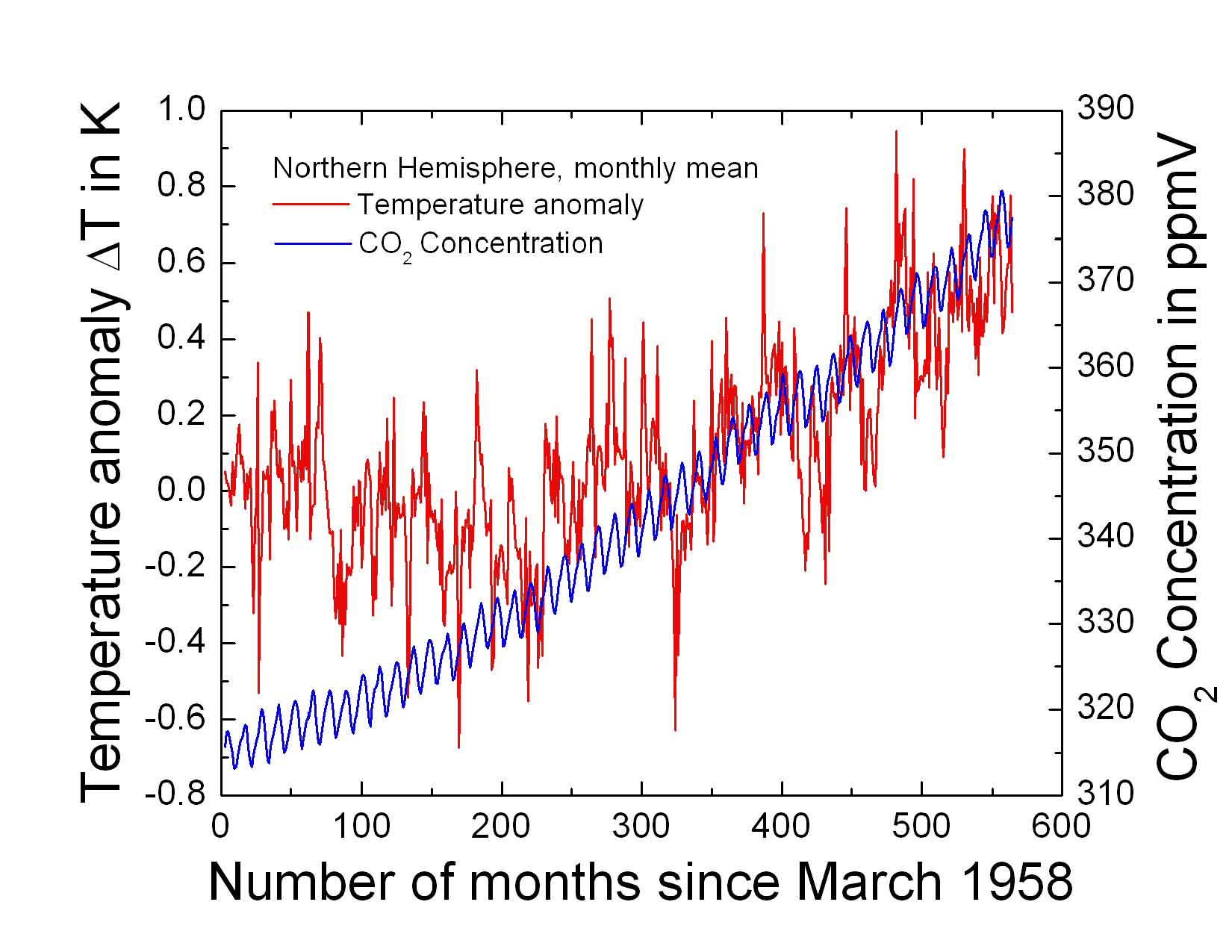}
\end{center}
\begin{center}
\includegraphics[width=.75\textwidth,height=!]{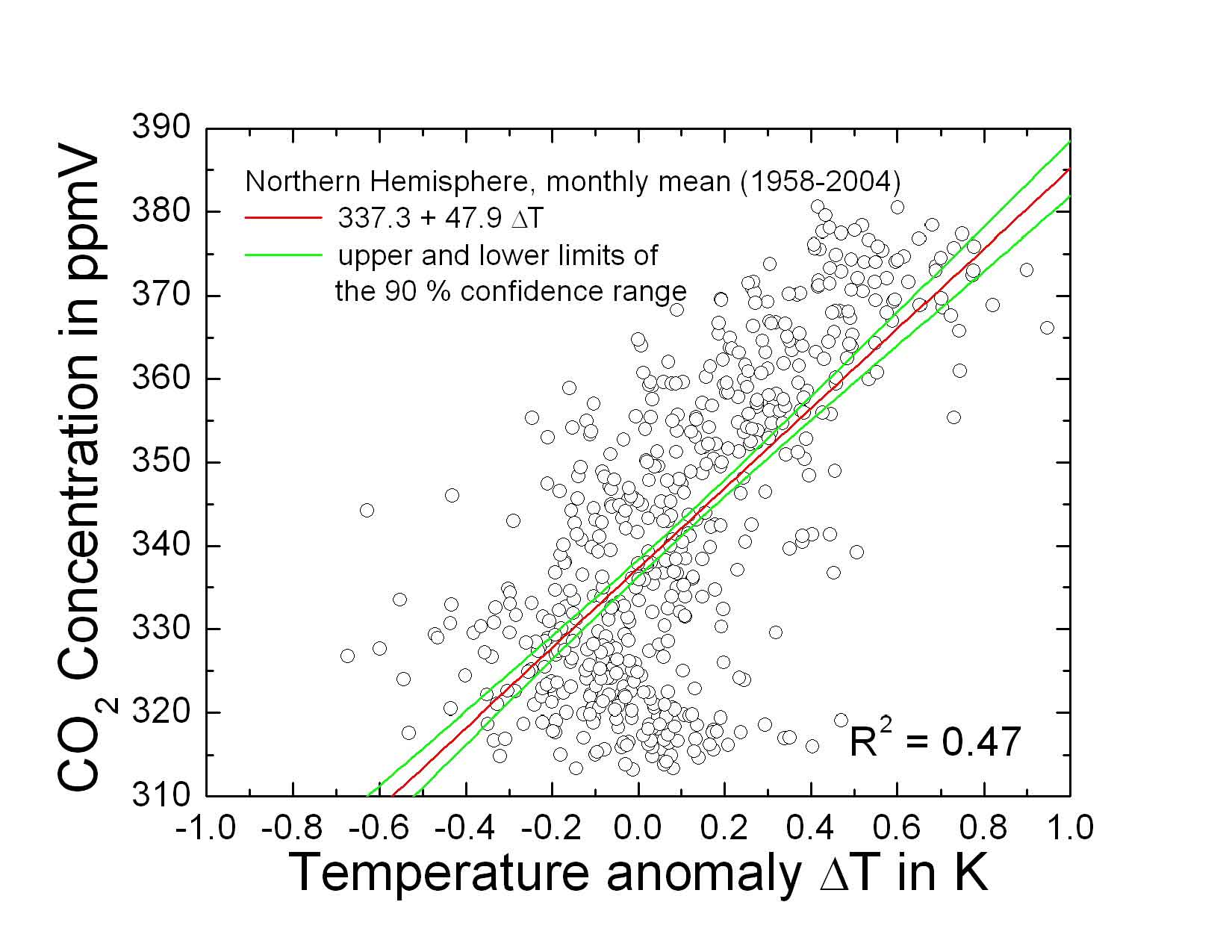}
\end{center}
\caption{Same as in Figure \ref{Figure_1}, but for the monthly mean
values.} \label{Figure_2}
\end{figure}
\begin{figure}[t]
\begin{center}
\includegraphics[width=.75\textwidth,height=!]{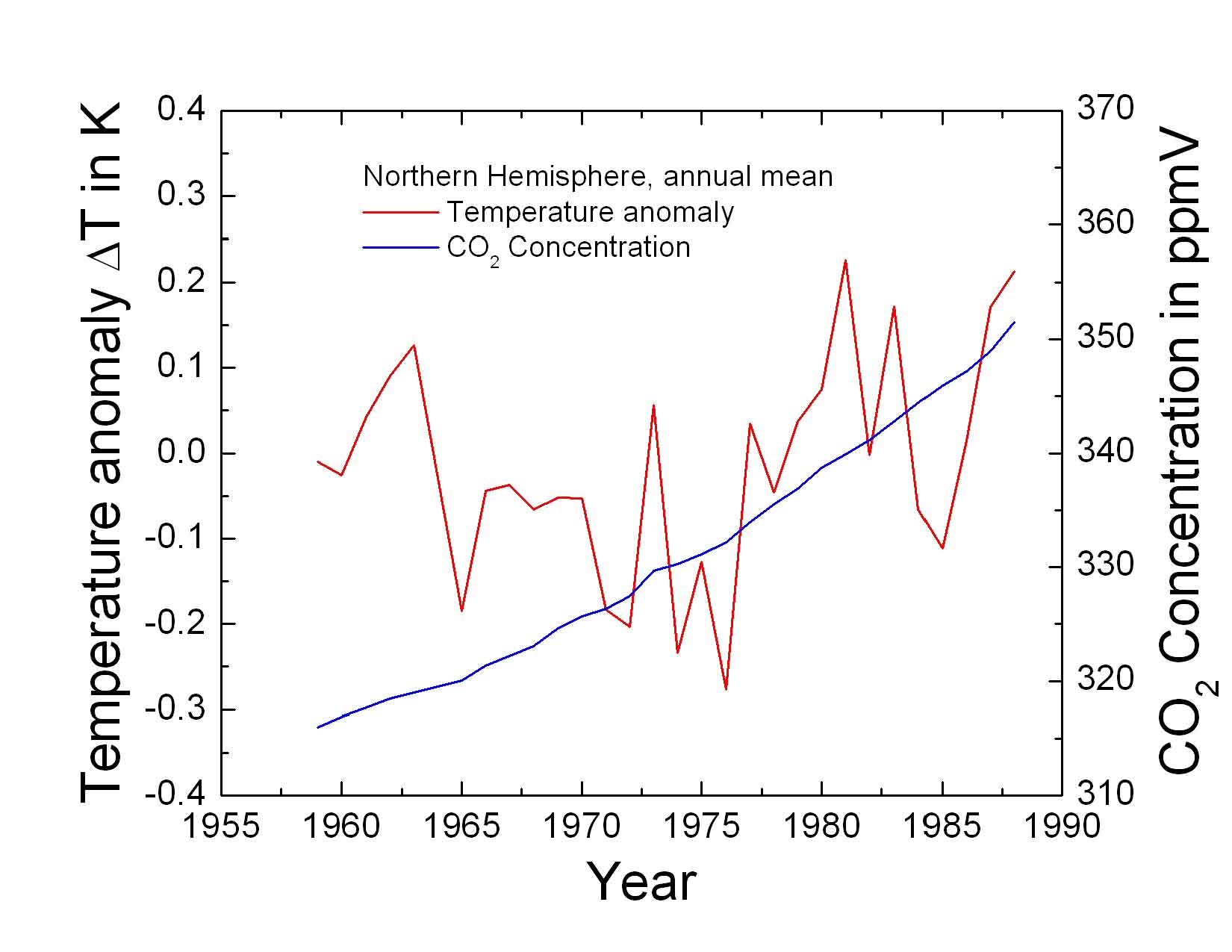}
\end{center}
\begin{center}
\includegraphics[width=.75\textwidth,height=!]{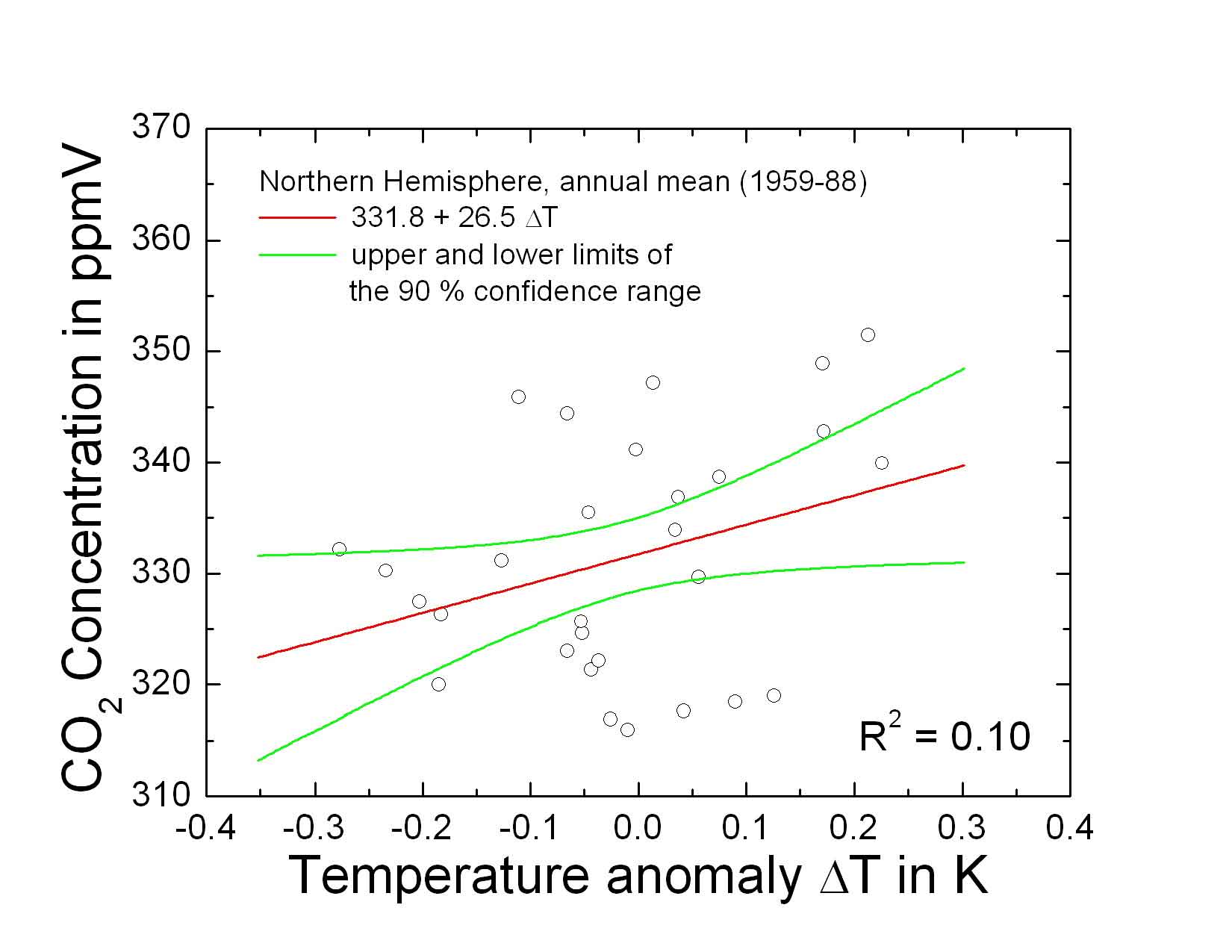}
\end{center}
\caption{Same as in Figure \ref{Figure_1}, but for the period 1958
to 1988} \label{Figure_3}
\end{figure}
\begin{figure}[t]
\begin{center}
\includegraphics[width=.75\textwidth,height=!]{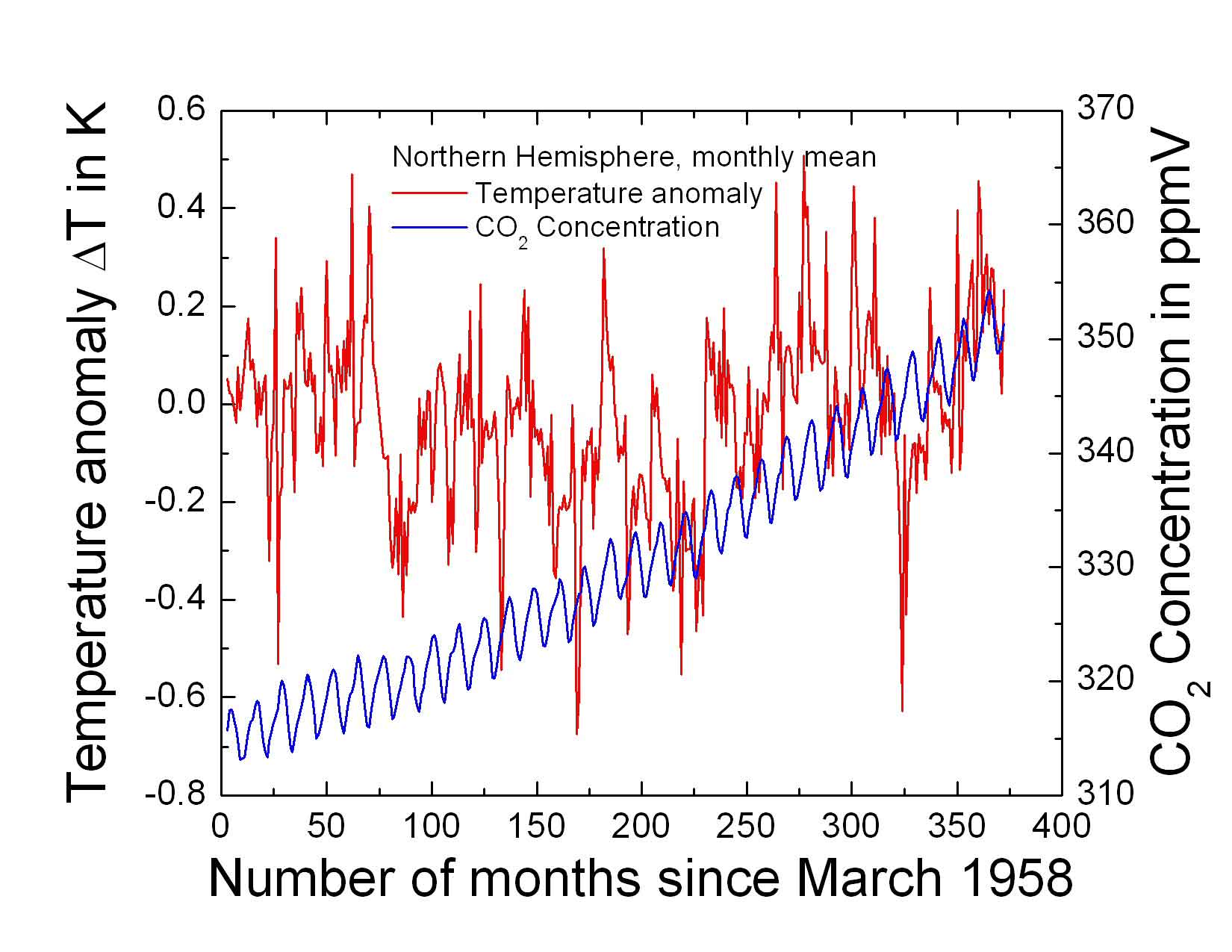}
\end{center}
\begin{center}
\includegraphics[width=.75\textwidth,height=!]{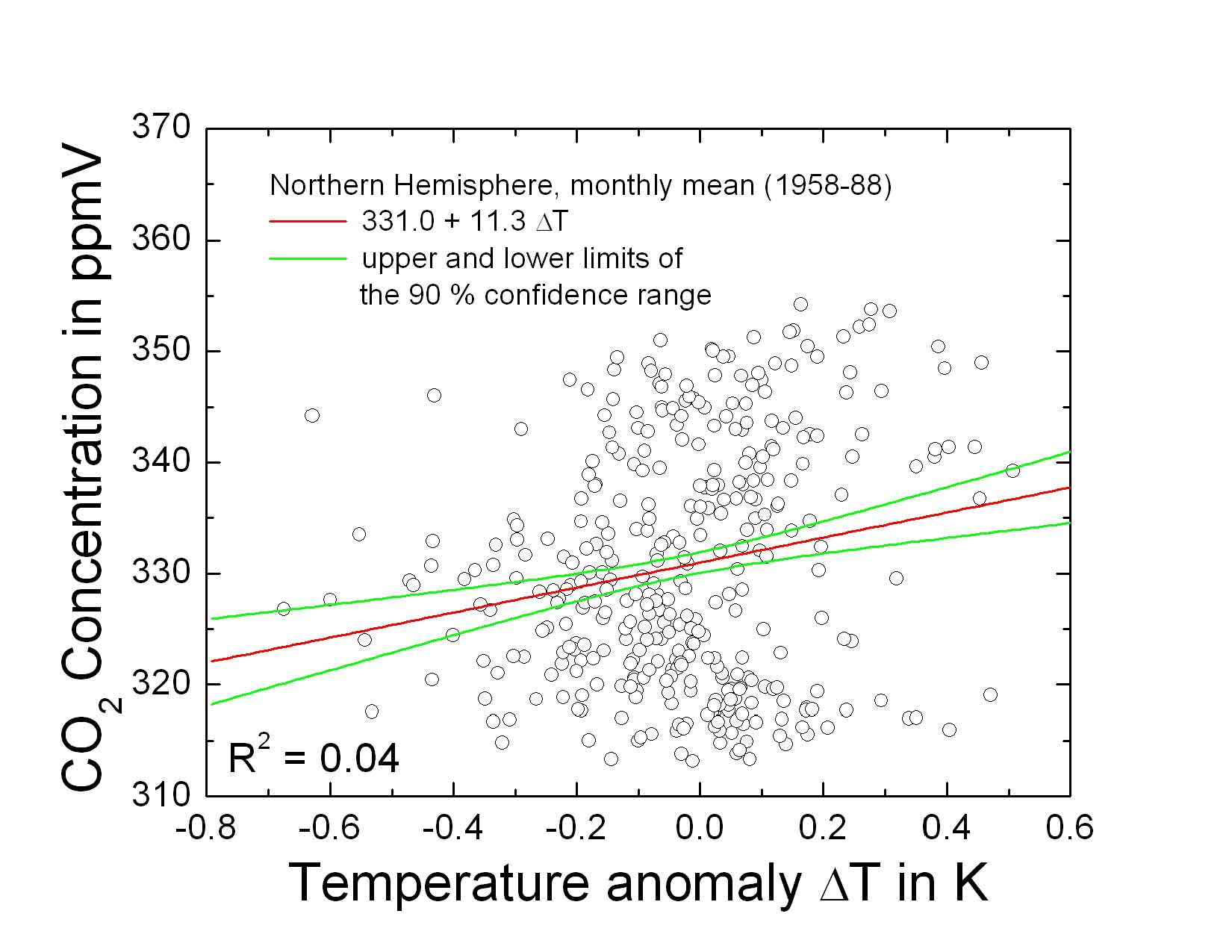}
\end{center}
\caption{Same as in Figure \ref{Figure_2}, but for the period 1958
to 1988.} \label{Figure_4}
\end{figure}

\pagestyle{myheadings} \markboth{\centerline{\textsc{Gerhard
Kramm}}}{\centerline{\textsc{Comment to “Recent Climate Observations
Compared to Projections”\dots}}}
\noindent With great interest I read this article of Rahmstorf et
al. \cite{Ra07}. It is surprising to me that the authors only
consider a period from the beginning of the seventies to recent
years. I think that this is, clearly, a source of
misinterpretation.\\ \indent The Mauna Loa observation of the
atmospheric carbon dioxide ($CO_2$) concentration (probably the best
$CO_2$ data we have) started in 1958. Therefore, one should consider
the whole period of these observations. As illustrated in Figures
\ref{Figure_1} and \ref{Figure_2}, the correlations for the period
1958 to 2004 show a somewhat different picture as presented by
Rahmstorf et al. \cite{Ra07} in their Figure 1. The results of my
figures are based on the Mauna Loa $CO_2$ data (monthly and annual
averages) and the mean near surface temperature anomalies of the
Hadley Centre for Climate Prediction and Research, MetOffice, UK,
for the northern hemisphere (also monthly and annual averages),
too.\\ \indent If we do not consider the whole period of available
data, then we might run in the wrong direction. Figures
\ref{Figure_3} and \ref{Figure_4}, for instance, illustrate results
from correlation calculations for the period ranging from 1958 to
1988. Remember that in 1988 the Intergovernmental Panel of Climate
Change (IPCC) of the United Nations and the World Meteorological
Organization (WMO) was established. As shown in the figures
attached, during 1988 there was certainly no correlation between
$CO_2$ and the temperature anomalies, neither on the annual time
scale (Figure \ref{Figure_3}) nor on the monthly time scale (Figure
\ref{Figure_4}). Consequently, I wonder why the IPCC was established
during that time.

\end{document}